\documentclass{PoS}

\title{``Oscillating'' components in the BL Lac object 0716+714?}

\ShortTitle{0716+714: oscillating jet components}

\author{\speaker{S. Britzen}$^a$, V.~Meyer$^a$, A.~Witzel$^a$, I. Agudo$^a$, M.F. Aller$^b$, H.D. Aller$^b$, A. Eckart$^c$, and J.A. Zensus$^a$\\
\llap{$^a$}Max-Planck-Institut f\"ur Radioastronomie, Auf dem H\"ugel 69, 53121 Bonn, Germany\\
\llap{$^b$}Astronomy Department, University of Michigan, Ann Arbor, MI 48109-1042, USA\\
\llap{$^c$}I. Physikalisches Institut K\"oln, Universit\"at zu K\"oln, Z\"ulpicher Str. 77, 50937 K\"oln, Germany\\
E-mail: \email{sbritzen@mpifr-bonn.mpg.de},
	        \email{vmeyer@mpifr-bonn.mpg.de},
	        \email{awitzel@mpifr-bonn.mpg.de},
	        \email{iagudo@mpifr-bonn.mpg.de},
		\email{mfa@umich.edu},
		\email{haller@umich.edu},
		\email{eckart@ph1.uni-koeln.de},
		\email{azensus@mpfir-bonn.mpg.de}}

\abstract{15 VLBA observations of the BL Lac object 0716+714, performed between 1994.67 and 2006.40 at 15 GHz, have been analyzed.  Part of the data (12 epochs) were obtained within the Mojave-survey project and reanalyzed by us.  We present a new motion scenario for jet model component motion in this source which suggests that no longterm-outward motion but rather an oscillation of components around an average core separation is taking place. Although no significant outward motion in the core separation can be found, motion with regard to the position angle is observed. We give lower limits for the derived apparent motions and Doppler factors and compare them with values published in the literature. We find a relation between the total flux-density evolution at 14.5 GHz (UMRAO data) and the position angle changes observed for the jet component closest to the core. We suggest a significant geometric contribution to the longterm flux-density variability in 0716+714.}

\FullConference{The 8th European VLBI Network Symposium on New Developments in VLBI Science and Technology
                and EVN Users Meeting  \\
		September 26-29 2006\\
		Torun, Poland}

\begin{document}
\section{Introduction}
The BL Lac Object 0716+714 (redshift $\geq$ 0.3, \cite{wagner}) shows flux-density variability at all wavelength ranges that have been investigated so far (e.g., \cite{wagnerwitzel}). It is still the only AGN for which a simultaneous change in the mode of the radio/optical variability has been observed on intraday time scales (\cite{quirrenbach}) and is thus so far the best candidate for an intrinsic origin of intraday variability (IDV, \cite{wagnerwitzel} and references therein).
 0716+714 has been investigated with VLBI  at various radio frequencies with different arrays. On parsec-scales it reveals a North-South directed jet with several components imbedded (e.g., \cite{eckart}). On kpc-scales, the jet is more East-West oriented with some evidence for halo-emission (e.g., \cite{antonucci}). A rather large range in proper motions (0.05 mas/yr - 1.1 mas/yr) has been reported by different authors based on the investigation of an increasing core separation with time modelled for jet components in this source (e.g., \cite{eckart}, \cite{witzel}, \cite{schalinski}, \cite{gabuzda}, \cite{jorstad}, \cite{torres}, \cite{kellermann}, \cite{bach}). Based on these results it is not clear, whether 0716+714 reveals the slow velocities typical for BL Lac objects (e.g., \cite{britzen1}) or faster velocities more typical for quasars (e.g., \cite{kellermann}). We here present first results of an (re-)analysis of 15 epochs of VLBA observations  revealing evidence for a new motion scenario in the jet of this source. In this paper, we concentrate on the description of the results based on the 15 GHz data. A paper describing these results in more detail and the implications of a multifrequency analysis is in preparation.
  In contrast to previous studies - that relied mainly on investigating the core separation evolution - we study the position angle evolution with time as well. Again in contrast to previous studies, based on our new identification scenario, we find no evidence for any longterm-outward motion but instead find that all modelled jet components remain at similar core separations and move with regard to the position angle. We call this kind of motion ``oscillatory''. A similar motion phenomenon has already been observed in the jet of 1803+784 (\cite{britzen1803}, \cite{nadia}, \cite{britzen18032}).

\section{The observations}
0716+714 has been observed in 15 VLBA observations at 15 GHz between 1994.67 and 2006.40. 12 of these images have been obtained by the MOJAVE / 2 cm-VLBA survey group (MOJAVE, e.g., \cite{lister}, 2cm Survey, e.g., \cite{kellermann}). 0716+714 was used as calibrator source for observations targeted on NRAO 150 (\cite{agudo1}{) in the epochs 2004.64, 2004.97, and 2005.25. Part of the data presented here have already been analyzed by \cite{bach}. We show the results of a reanalysis of all the data sets. Modelfits within the difmap package (e.g., \cite{lovell}) have been performed starting from a point-like model in all the observations.
The final number of jet components necessary to fit the data adequately was reached when adding another jet component did not lead to a significant improvement of the value of chi-square.
The uncertainties of the model component parameters have been determined by comparing the parameter ranges obtained by performing modelfits with different numbers of model components ($\pm$1 component). 
\subsection{Component identification}
The component identification was done based on the assumptions that changes of the flux-density, the core separation, and the position angle of the modelled jet components should be small on time scales between adjacent epochs. We show an example for the component identification in Fig.\ref{f1}~(left). This identification across the epochs leads to a new motion scenario (described in the next section) for this source compared with earlier scenarios published in the literature. 
\begin{figure}[htb]
\begin{center}
\hspace*{-0.6cm}\includegraphics[clip,width=8.2cm,angle=-90]{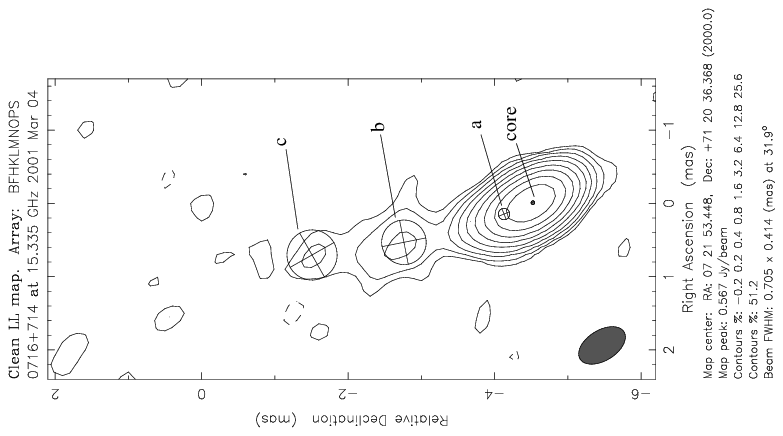}
\hspace*{0.5cm}\includegraphics[clip,width=7.8cm,angle=-90]{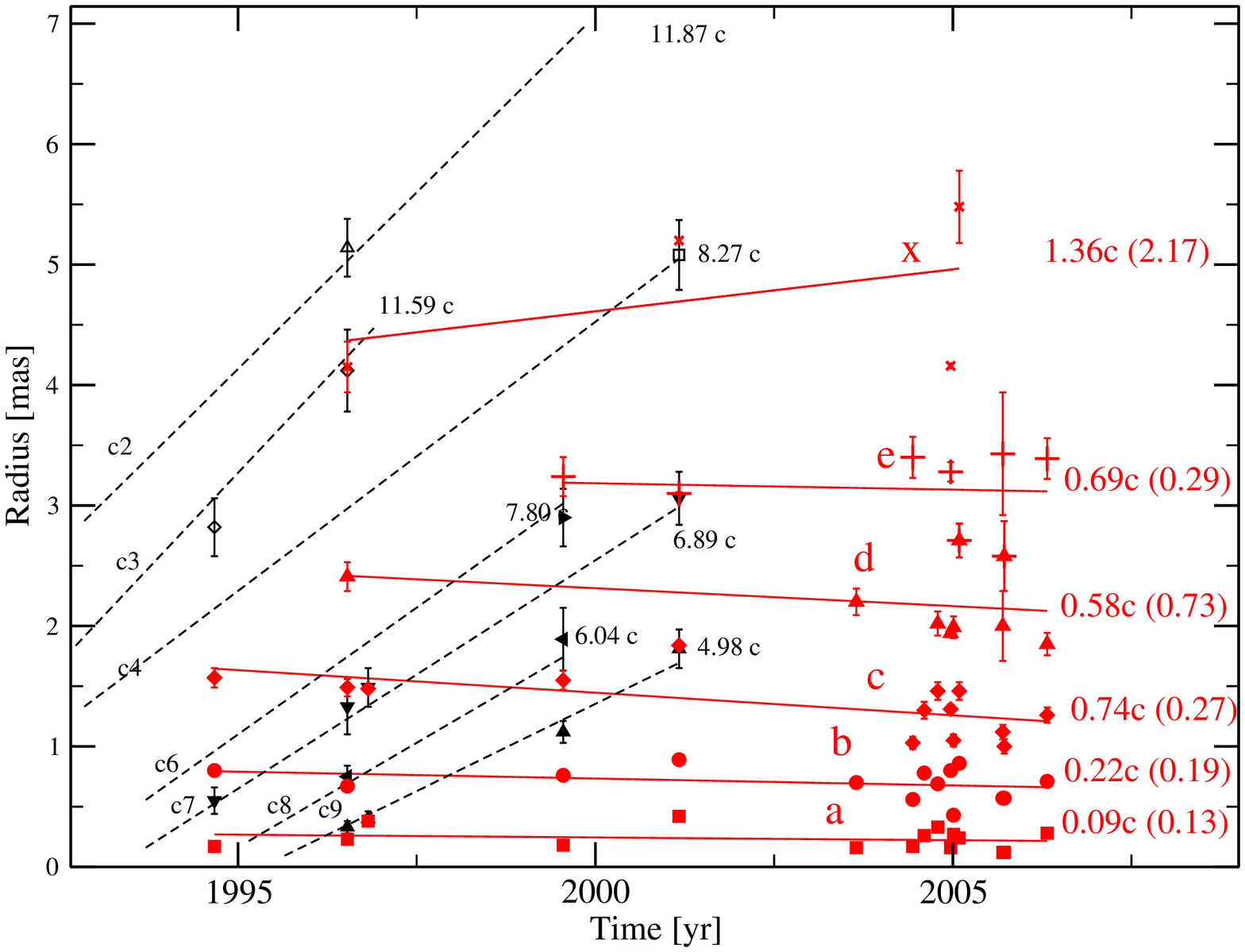}
\end{center}
\caption{We show the source structure in a map obtained at 15 GHz with the contours of the model superimposed (left). The inner three jet components are labelled. The core separation as function of time based on the 15 GHz data is shown in the figure on the right. A  linear regression based on a previous identification scenario (for a subsample of the data) is shown with a dashed line (components {\bf c2}-{\bf c12}, \cite{bach}). Components {\bf a}-{\bf x} result from the new modelfits (red). The linear regression for this new identification is plotted as solid line. The absolute values of the apparent motion with uncertainties in brackets are listed.}
\label{f1}
\end{figure}

\section{Results}
\subsection{Oscillatory components}
In Fig.\ref{f1}~(right) we show the core separation as a function of time for all model-fitted components from the 15 GHz data. In addition to our own results, the results published by \cite{bach} for a subset of the data presented here are shown as well. The component identification across the observing epochs by Bach et al. is based on a combined data set at 5, 8, 15 and 22 GHz. For a better comparison with our component identification, we here only show the 15 GHz data with the linear regression made by \cite{bach} (dashed line) superimposed. The larger amount of data analyzed here leads to a different component identification (solid line). The fast apparent velocities found by \cite{bach} (noted in Fig.\ref{f1}~(right)) can not be confirmed within our new identification of model components. Instead, we find that our derived values for the apparent proper motions of model components are consistent with no longterm outward-motion if we calculate these values based on the core separation versus time.
All jet components show significant changes with regard to the position angle. We show this relation for the inner three jet components in Fig.\ref{f2}~(left). In Fig.\ref{f2}~(right) we show the motion of these model components in rectangular coordinates. 
\begin{figure}[htb]
\begin{center}
\hspace*{-1.9cm}\includegraphics[clip,width=5.6cm,angle=-90]{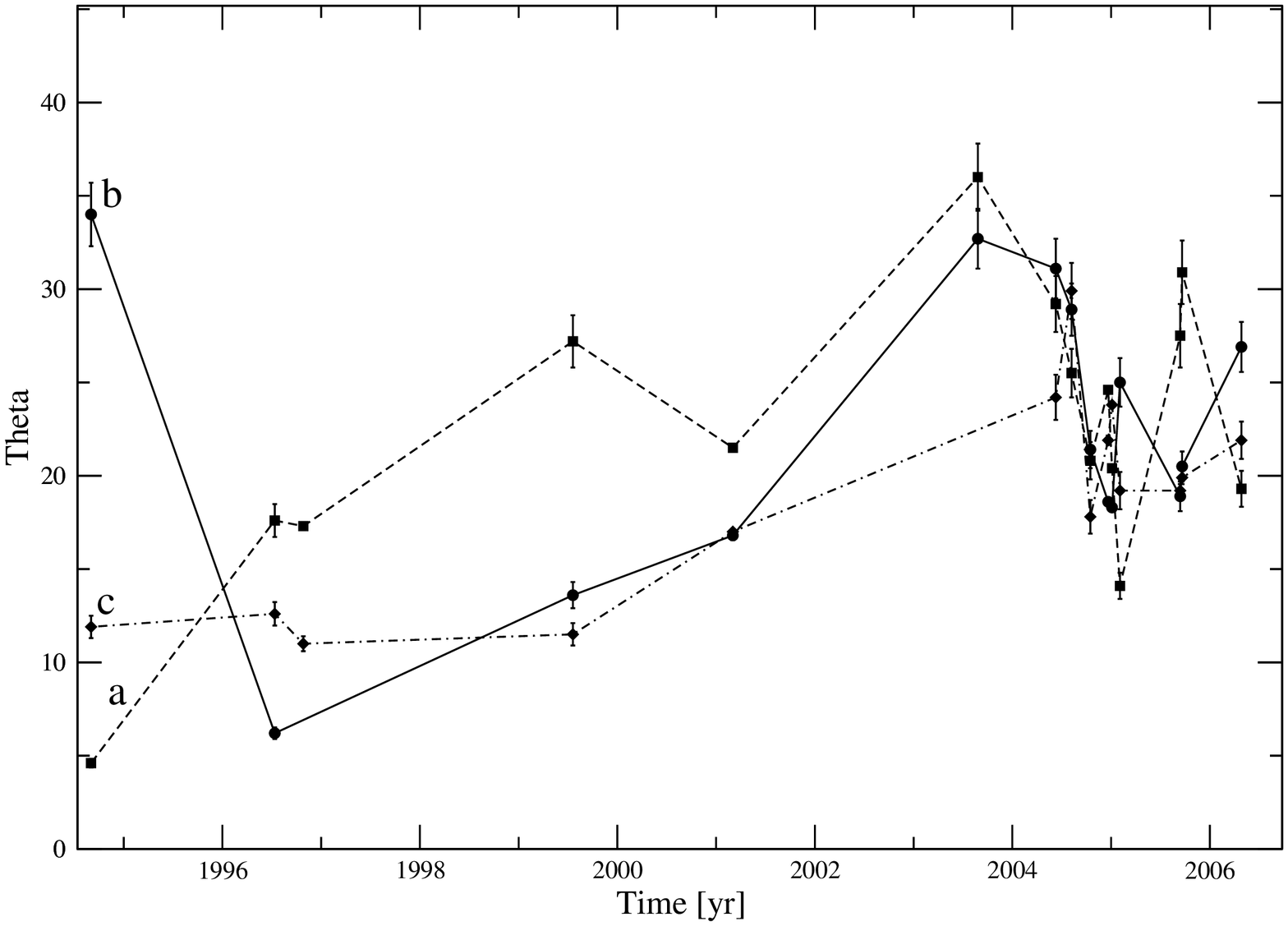}
\hspace*{1.5cm}\includegraphics[clip,width=5.6cm,angle=-90]{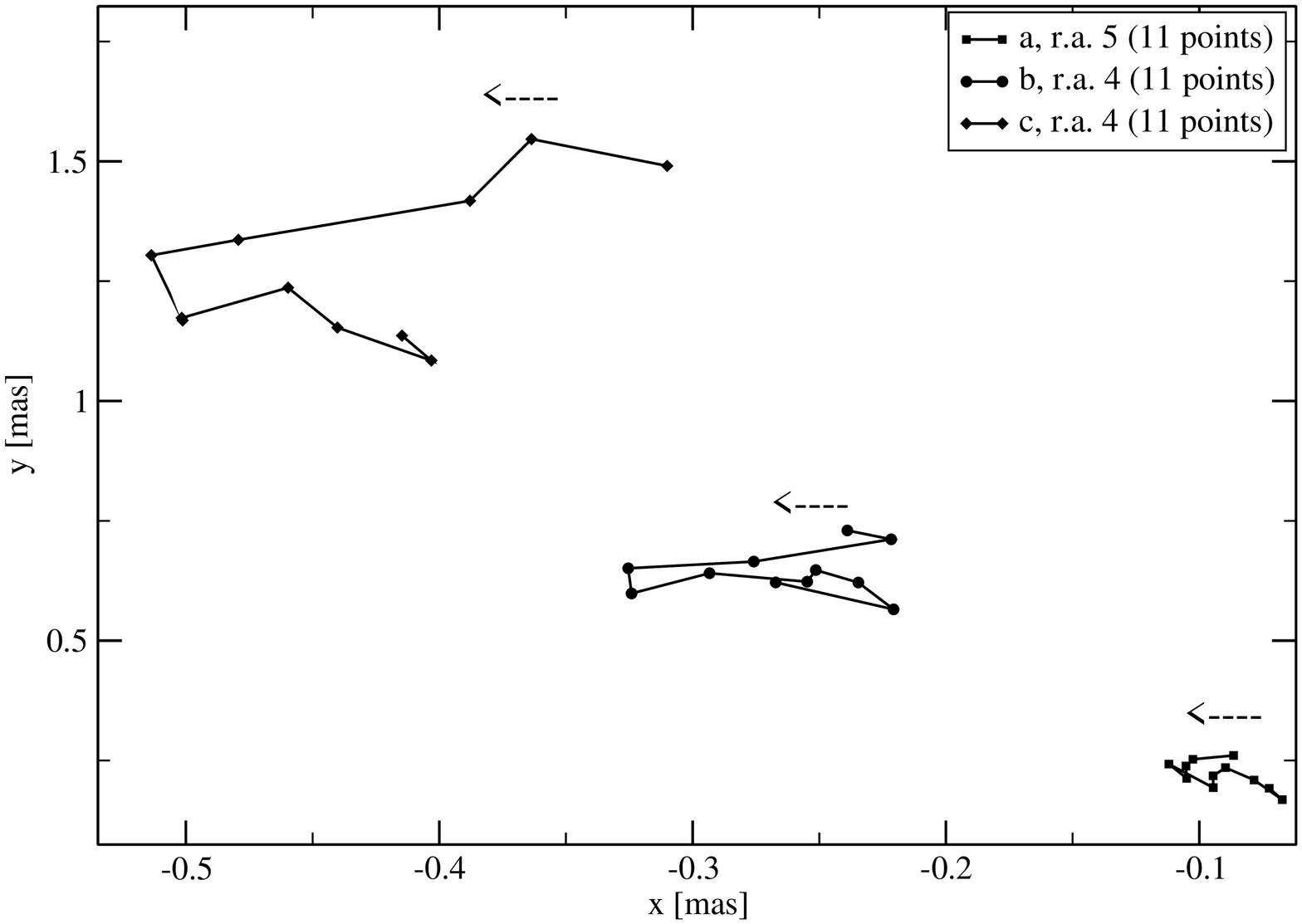}
\end{center}
\caption{The position angle as function of time is shown for the inner three model components ({\bf a} - {\bf c}) (left). On the right, the inner component's motion are shown in rectangular coordinates (5-point running average for {\bf a}, 4-point running averages for {\bf b} and {\bf c}).}
\label{f2}
\end{figure}

\subsection{Apparent velocities and Doppler factors}
Some of the model components reveal apparent inward motion, which is not uncommon in AGN. However, none of these values is of 3 sigma significance.
For the apparent velocities derived for the motion based on the core separation evolution with time only, we find values between 0.1$\pm$0.1$c$ and 1.4$\pm$2.2$c$. The Doppler factors for this motion are between 1.1$\pm$1.2 and 3.0$\pm$4.4 depending on the assumed angle to the line of sight of 4.9$^\circ$ or $\simeq$0$^\circ$, respectively (angles to the line of sight taken from \cite{bach}). For the non-radial motion, we derive preliminary lower limits for the apparent velocities between 5$\pm$3$c$ and 10$\pm$5$c$ and Doppler factors between 8$\pm$4 (inner component {\bf a}) and 11$\pm$5 (outer component {\bf d}) (4.9$^\circ$) and 9$\pm$5 ({\bf a}) and 20$\pm$10 ({\bf d}) ($\simeq$ 0$^\circ$) for the individual components. These Doppler factors have been derived taken the motion in rectangular coordinates into account. They compare well with Doppler factors derived based on e.g., Inter-Day variability (\cite{agudo1}, \cite{fuhrmann}). 
 \subsection{Correlation between total flux-density evolution \& position angle evolution}
In Fig.\ref{f3} we show a composition of several data sets: the total flux density at 14.5 GHz (black dots, UMRAO data), the flux-density of the core (dot-dashed), the flux-density of component {\bf a} (closest component to the core, dashed), and the position angle for component {\bf a} (dotted) as function of time.We find that most of the total flux-density is contained in the core component. We also find evidence for a positive relation between the total flux-density evolution with time and the position angle evolution of component {\bf a} with time. This might suggest that the longterm flux-density evolution in 0716+714 may have a non-negligible geometric contribution. 

\begin{figure}[htb]
\begin{center}
\hspace*{0.5cm}\includegraphics[clip,width=9.2cm,angle=-90]{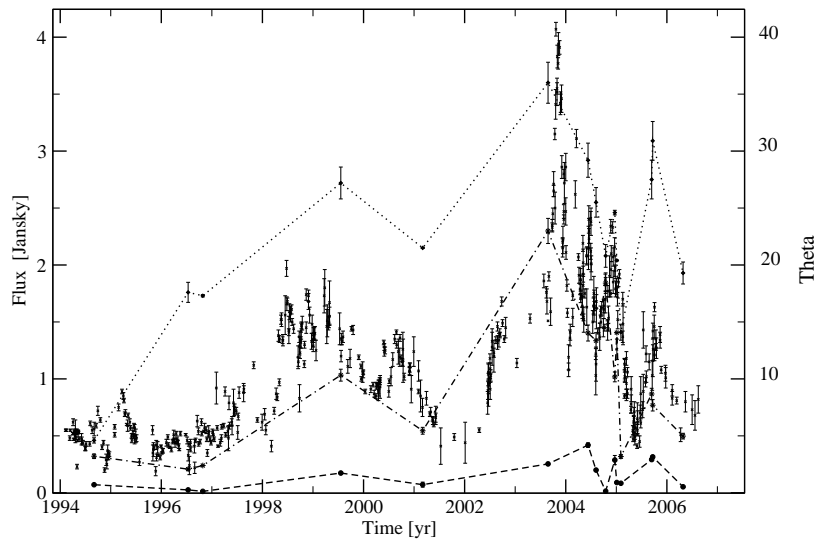}
\end{center}
\caption{The total flux-density at 14.5 GHz (UMRAO data) is shown as a function of time. The VLBI core flux-density (15 GHz, dot-dashed line) and the VLBI flux-density of component {\bf a} (dashed line) are superimposed. The left y-axis measures the flux-density. In addition, the position angle for component {\bf a} (component closest to the core) is shown as a function of time (dotted line). The right y-axis denotes for the position angle.
}
\label{f3}
\end{figure}

\section{Conclusions}
We present a new model component motion scenario for 0716+714 with the following implications:\\
$\bullet$ In this scenario the jet components show no 3 sigma evidence for longterm-outward motion, instead they remain at similar core separations and seem to change significantly in position angle within our computed uncertainties.\\
$\bullet$ Preliminary lower limits for the apparent velocities (with regard to the position angle motion) lie between 5$c$ (4.9$^\circ$) and 10$c$ ($\simeq$0$^\circ$), the Doppler-factors between 8-11 (4.9$^\circ$) and 9-20 ($\simeq$0$^\circ$) for the inner to outer components. These values compare well with the limits derived e.g., from variability arguments.\\
$\bullet$ We find a relation between the total flux-density and the position angle of the component closest to the core ({\bf a}), which might suggest a non negligible geometric origin of the longterm flux-density variability in 0716+714.\\
A model to explain the origin of this correlation (based on rotation or precession) is currently in preparation. In addition, we are investigating a large set of multifrequency data (5, 8, 22, 43 GHz) to study the frequency dependence of the presented relations.\\
The usually observed outward motion of jet components in AGN is - according to our analysis - not observable in the BL Lac object 0716+714. Since BL Lac objects most likely constitute the class of AGN with a jet directed almost in the direction of the line of sight, projection effects, e.g. the amplification of curved structures due to the small angle to the line of sight, might play an important role and will have to be disentangled before we draw final conclusions.\\

\vspace*{0.3cm}
\noindent
{\bf Acknowledgements:}\\
I. Agudo acknowledges financial support from the EU Comission for Science and Research under contract HPRN-CT-2002-00321 (ENIGMA Network).
This research has made use of data from the University of Michigan Radio Astronomy Observatory which has been supported by the University of Michigan and the National Science Foundation.
We also aknowledge the MOJAVE and VLBA 2-cm Survey Program teams.

\end{document}